\documentclass[journal]{IEEEtran}  

\IEEEoverridecommandlockouts                              

\usepackage{upquote}
\usepackage{amsmath,bm}
\usepackage[pdftex]{graphicx}
\usepackage[compress]{cite}
\usepackage{mathtools}
\usepackage{accents}
\usepackage{cases}
\usepackage{amsmath}
\usepackage{empheq}
\usepackage{amssymb}
\usepackage{physics}
\usepackage{breqn}

\usepackage[ruled,norelsize]{algorithm2e}
\usepackage{todonotes}
\usepackage{booktabs,chemformula}
\DeclareMathOperator*{\argmaxA}{arg\,max}
\DeclareMathOperator*{\argminA}{arg\,min}
\DeclareMathOperator{\E}{\mathbb{E}}
\newtheorem{assumption}{Assumption}
\newtheorem{lemma}{Lemma}
\newtheorem{theorem}{Theorem}
\newtheorem{corollary}{Corollary}
\newtheorem{remark}{Remark}
\newtheorem{proposition}{Proposition}

\newcommand{\M}{\mathcal{M}}

\NewDocumentCommand{\evalat}{sO{\big}mm}{%
  \IfBooleanTF{#1}
   {\mleft. #3 \mright|_{#4}}
   {#3#2|_{#4}}%
}

\newcommand{\expect}{\operatorname{\E}\expectarg}
\DeclarePairedDelimiterX{\expectarg}[1]{[}{]}{%
  \ifnum\currentgrouptype=16 \else\begingroup\fi
  \activatebar#1
  \ifnum\currentgrouptype=16 \else\endgroup\fi
}

\newcommand{\innermid}{\nonscript\;\delimsize\vert\nonscript\;}
\newcommand{\activatebar}{%
  \begingroup\lccode`\~=`\|
  \lowercase{\endgroup\let~}\innermid 
  \mathcode`|=\string"8000
}

\ifodd 1

\newcommand{\com}[1]{{\color{red}{Comment: #1}}}
\else

\newcommand{\com}[1]{}
\fi

\title{Distributed Estimation in the Presence of Strategic Data Sources}

\author{Kewei Chen, Donya Ghavidel, Vijay
  Gupta, and Yih-Fang Huang
  \thanks{The authors are with the Department of Electrical Engineering, University of Notre Dame, IN 46556 {\tt \small \{kchen6,dghavide, vgupta2,huang\}@nd.edu}. Research is supported by NSF CNS-1739295, NSF CNS-1544724, NSF ECCS-1550016 and ARO W911NF-17-1-0072. A preliminary version of the formulation here was presented in the IEEE Conference on Decision and Control (CDC) 2016. Almost all the results here are new as compared to that.}}

\begin{document}
\bstctlcite{IEEEexample:BSTcontrol}
\maketitle
\thispagestyle{empty}
\pagestyle{empty}

\begin{abstract}
Distributed estimation that recruits potentially large groups of humans to collect data about a phenomenon of interest has emerged as a paradigm applicable to a broad range of detection and estimation tasks. However, it also presents a number of challenges especially with regard to user participation and data quality, since the data resources may be strategic human agents instead of physical sensors. We consider a static estimation problem in which an estimator collects data from self-interested agents. Since it incurs cost to participate, mechanisms to incentivize the agents to collect and transmit data of desired quality are needed. Agents are strategic in the sense that they can take measurement with different levels of accuracy by expending different levels of effort. They may also misreport their information in order to obtain greater compensation, if possible. With both the measurements from the agents and their accuracy unknown to the estimator, we design incentive mechanisms that encourage desired behavior from strategic agents. Specifically, we solve an optimization problem at the estimator which minimizes the expected total compensation to the agents while guaranteeing a specified quality of the global estimate. 
\end{abstract}


\section{Introduction}
Distributed estimation theory to solve the problem of fusing data from a group of sensors to estimate a parameter or a random variable is a well-developed field. More recently, the emerging areas of social computing and crowdsourcing have enabled many large scale sensing and estimation tasks that leverage many humans (or human owned and operated devices) to collect data about phenomena of interest (see works such as~\cite{guo2014participatory,ma2014opportunities} for an overview). An example is that of aggregating information and opinions of a `crowd' recruited using Amazon Mechanical Turk to perform tasks that are time consuming and difficult to scale such as image labeling. Similar applications have been proposed or demonstrated in the fields ranging from environmental monitoring \cite{boulos2011crowdsourcing}, health data collection \cite{pryss2015mobile}, traffic monitoring \cite{pan2013crowd}, and so on. 

Beyond already existing challenges in traditional distributed estimation or detection~\cite{berthet2016resource,cattivelli2010diffusion,varshney2012distributed}, new challenges arise in the design of such a crowdsensing system since data sources may not have any incentive to provide the data aggregator with the quality of data that it desires~\cite{allahbakhsh2013quality}. This might be due to the fact that sensors may have to exert resources (e.g., time, power, or bandwidth) to produce an accurate measurement~\cite{gao2015survey}. Further, even though the sensors may have accurate data, they may still wish to corrupt data before transmission either to gain privacy or for some other selfish reason~\cite{gong2016optimal}. Early work in this field (e.g., \cite{mohan2008nericell,thiagarajan2009vtrack,rana2010ear}) ignored these issues and assumed that participants were voluntary recruits who would collect and provide high quality data. More recently, it has been recognized that without a suitable incentive being present, such voluntary providers of data may not be enough to generate an estimate of desired quality. As an illustrative example, \cite{hu2006can} studied product reviews on Amazon.com and concluded that users with a moderate outlook are unlikely to report; thus, while controlled experiments on the same items reveal normally distributed opinions, voluntarily reported ratings often follow bi-modal, U-shaped distributions where most of the ratings are either very good or very bad. 

Accordingly, there has been recent work on designing mechanisms that incentivize data sources (i) to participate and generate measurements of sufficient quality (i.e. effort exertion), and (ii) to report these measurements and their quality accurately (i.e. truthful elicitation). Incentivization may be through monetary or non-monetary rewards for the sensors. A review of various incentive mechanisms, including both monetary and non-monetary incentives, is provided in \cite{gao2015survey,jaimes2015survey,restuccia2016incentive}. As an illustration, if the agents cannot falsify data and the problem is solely to incentivize effort exertion, mechanisms such as those in~\cite{cai2015optimum,farokhi2015budget, witkowski2013dwelling, radanovic2016incentives, lee2010sell,cao2015target, chen2016auction} have been proposed. 
Similarly, for the problem of truthful elicitation, the class of mechanisms called peer prediction mechanisms has been developed with different information structures (see e.g., \cite{prelec2004bayesian, miller2005eliciting, jurca2009mechanisms, witkowski2012robust,radanovic2013robust}) to incentivize the agents to report truthfully in a game-theoretic equilibrium.

However, most of the works in the literature focused on either  truthfulness elicitation or effort elicitation, without considering how much total reward is to be paid, or the trade-off between the total payment and the estimate accuracy at the estimator. A systematic theory that addresses the challenges of incentive mechanism design with the objective of optimizing the overall cost function at the estimator is not well studied.
In this paper, we address the mechanism design problem at the estimator of minimizing expected total compensation to be made to the strategic agents while guaranteeing a specified quality of global estimate, with both the measurements and their accuracy from the strategic agents unknown to the estimator. 

The works that are closest seem to be \cite{luo2018parametric, liu2016learning,liu2017sequential,dobakh}. 
\cite{luo2018parametric} considered a model that determines the compensation to strategic agents by verifying their reports with the ground truth of the phenomenon of interest, which was assumed available to the estimator after the estimation process. However, this is not applicable in many cases where the reports to be collected are subjective, such as collecting ratings for a product from consumers. In this paper, we do not require the availability of ground truth.
\cite{liu2016learning} considered the mechanism design problem from the perspective of an estimator who cares about both the estimate accuracy and the total payment with the assumption that the actual effort costs of the agents are drawn from a known distribution. Although this assumption can be relaxed by learning the distribution in a sequential setting \cite{liu2017sequential}, their model is limited to a binary-answer task (e.g., an image contains a certain object or not) and a binary effort model (e.g., either exert a fixed level of effort at a fixed cost or no effort and no cost at all). 
Different from solving the discrete value tasks (i.e., detection problem or classification problem) such as image labeling and spectrum occupancy sensing, we focus on estimation tasks with continuous measurements and continuous effort models.
Finally, unlike \cite{dobakh} where the model is limited to a specific effort cost function, we do not restrict the format of effort cost function. 

The rest of the paper is organized as follows.
In Section II, the problem statement is presented. In Section III, we solve the optimization problem at the estimator to minimize total compensation while guaranteeing a certain estimation accuracy. Next, in Section IV, we present an optimal mechanism that achieves the desired behavior from the strategic agents in a Nash equilibrium when the cost functions of the strategic agents satisfy a certain property. In Section V, we provide a feasibility-guaranteed sub-optimal mechanism when the cost functions of the strategic agents do not satisfy that property. Section VI provides some simulation experiments and Section VII concludes the paper.

\paragraph*{Notation} 
$\E_X [f]$ denotes the expectation of function $f$ taken with respect
to the random variable $X$; when $X$ is explained from the context, the notation is abbreviated as $\E[f]$.
A Gaussian distribution is denoted by $\mathcal{N} (m, \sigma^{2})$ where $m$ is the mean and $\sigma$ is the standard deviation.
A tuple of $n$ elements is denoted with parentheses by $(e_1, e_2, \cdots, e_n)$.

\section{Problem Statement}   \label{sec_ch2_problem statement}
\paragraph*{Estimation Setup} Consider a scalar-valued random variable $X$ that is distributed according to a prior distribution $X \sim\mathcal{N} (0, \sigma^{2}_{x})$ and takes an unknown value $x$ in an experiment. An estimator (also called an aggregator) seeks to estimate the value $x$ using observations from $N$ sensors (also called agents). The $i$-th sensor generates an observation $y_{i}\in\mathbb{R}$ according to the relation
\begin{equation}
\begin{split}
y_{i} &= x + v_{i}, \\
\end{split}
\end{equation}
where $v_{i}$ is the measurement noise that is distributed according to a Gaussian distribution with mean 0 and variance $\sigma^{2}_{i}.$ We assume that the measurement noises for the $N$ sensors are mutually independent and further independent with the variable $X$. For notational ease, we denote by $\xi$ the reciprocal of the variance, i.e., $\xi_{x}=\sigma_{x}^{-2}$ and $\xi_{i}=\sigma_{i}^{-2}$.

For each sensor $i$, given the measurement $y_i$, the minimum mean square error (MMSE) estimate $\hat{x}_{i}$ and the corresponding local mean squared error (MSE) $\Sigma_{i}$ can be computed as  
\begin{equation}  \label{ch2_eq_local_estimate}
\hat{x}_{i}=\frac{\sigma^{2}_{x}}{\sigma^{2}_{x}+\sigma^{2}_{i}}y_{i},
\end{equation} 
\begin{equation} \label{ch2_eq_local_MSE}
\Sigma_{i}=\frac{1}{\sigma^{-2}_{x}+\sigma^{-2}_{i}}=\frac{1}{\xi_{x}+\xi_{i}}.
\end{equation}
We denote $\hat{x}_{i}$ as the local estimate and $\Sigma_{i}$ as the local MSE at the $i$-th sensor since these quantities are obtained based on the information at each sensor.
These local estimates can be fused to obtain the global MMSE estimate $\hat{x}_{g}$ 
using the relation~\cite{xu2014distributed}
\begin{equation}  \label{eq:fuse}
\Sigma_{g}^{-1}\hat{x}_{g} = \sum_{i=1}^{N}\Sigma_{i}^{-1}\hat{x}_{i},
\end{equation}
where $\Sigma_{g}$ is the global MSE corresponding to $\hat{x}_{g}$ and can be calculated as
\begin{equation}  \label{global MSE}
\Sigma_{g}^{-1} = \sum_{i=1}^{N}\Sigma_{i}^{-1} - (N-1)\sigma_{x}^{-2}= 
\xi_{x}+\sum_{i=1}^{N}\xi_{i}.
\end{equation}

\paragraph*{Effort Cost} The variance $\sigma_{i}^{2}$ affects the quality of the measurement at sensor $i$ and is assumed to be a parameter that is under the control of the sensor. In other words, the sensor can put in more {\em effort} and decrease the variance $\sigma_{i}^{2}$ while incurring a higher effort cost. The effort cost may represent usage of battery, time, or some other resource. For simplicity and without loss of generality, we assume that $\xi_{i}$ is the effort level of agent $i$ that incurs an effort cost $c_{i}(\xi_{i})$. We make some weak assumptions on the cost function that describes the effort cost.
\begin{assumption}
The cost function of each sensor $c_{i}(\xi_{i})$ satisfies the following properties:
\begin{itemize}
    \item $c_{i}(\xi_{i}) \geq 0$, i.e., effort cost is non-negative;
    \item ${\dfrac{\partial c_{i}(\xi_{i})}{\partial \xi_{i}}} > 0$, i.e., more effort cost is incurred to obtain a measurement with higher accuracy;
    \item $\xi_{i} \in [0, \xi_{iu}]$ and $c_{i}(\xi_{i}) \in [0, c_{iu}]$.
\end{itemize}
\end{assumption}
Note that when sensor $i$ does not put in any effort, i.e., $\xi_i=0$ and $\sigma^2_i = \infty$, then the effort cost is zero, i.e., $c_i(0)=0$ and its local MSE is equal to the variance of the prior distribution of $X$, i.e., $\Sigma_i = \sigma^2_x$. 

\paragraph*{Formulation as a Mechanism Design Problem} We are interested in a formulation in which the estimator and the sensors are all self-interested. The estimator is interested in generating a global estimate with a specified accuracy as measured by the global MSE. To do so, it must incentivize sensors to generate and transmit measurements with sufficiently low local MSE. On the other hand, the sensors do not gain directly from the estimator being able to generate an accurate global estimate. Since they incur effort costs to generate measurements with low local MSE, the estimator must compensate the sensors using a payment mechanism of some sort, for simplicity, we assume the payment is monetary, although money may be thought of as a proxy of some other resource such as battery charging. The problem we consider in this paper is to minimize the payment from the estimator to incentivize self-interested sensors to obtain and report measurements with sufficient accuracy that allow the global MSE to be below a specified level.

\begin{figure}[tpb]
  \begin{center}
    \centerline{\includegraphics[width=\linewidth]{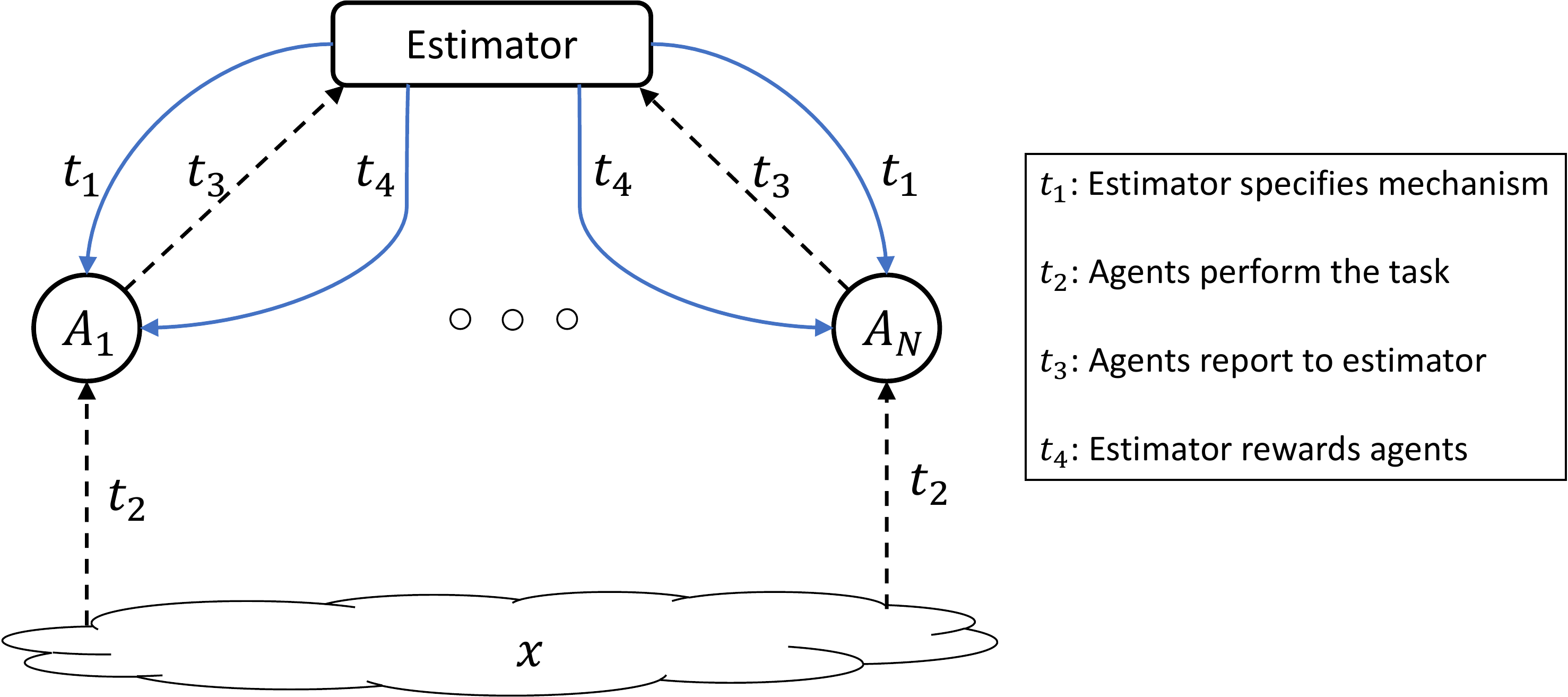}}
    \caption{Timeline and communication topology of the incentive mechanism design problem.}
    \label{fig/ch2_communication topology}
  \end{center}
\end{figure}

We now formulate this interaction as a mechanism design problem. The timeline of the interaction is as shown in Fig~\ref{fig/ch2_communication topology}. 
The estimator asks each sensor to report its measurement and local estimate. Note that reporting this pair is equivalent to reporting the local estimate and the local MSE. The strategy sets and the utility functions of each player are given as below.
\begin{itemize}
\item {\em Strategy sets:} Each sensor can choose the level of effort to exert and the values of its measurement and local estimate that it reports. For each sensor $i$, we define its strategy as choosing each element in the following tuple
\begin{equation*}
    s_{i} = (\xi_i, \hat{x}_{ri}, y_{ri}), 
\end{equation*}
where $\hat{x}_{ri}$ is the reported local estimate and $y_{ri}$ is the reported measurement. 
Denote the set of all feasible $s_i$'s by $S_i$. With a slight abuse of standard notation in game theory, when sensor $i$ adopts strategy $s_{i}$, denote by $s_{-i}=(s_1, s_2, \cdots, s_{i-1}, s_{i+1}, \cdots, s_N)$ the strategy profile of all the other sensors except for sensor $i$. The estimator decides how much payment each sensor $i$ will obtain and how to fuse the reports from the sensors. Since the sensors may misreport their local estimates,  (\ref{eq:fuse}) may not be the optimal way to fuse local reported estimates from the sensors.     
Thus, the strategy of the estimator includes the payment functions that map each strategy profile of the sensors to their payments and the fusion rule, i.e.,
\begin{equation*}
    s_{e} = (p_{i}(s_{1},\cdots,s_{N}),\ell(s_{1},\cdots,s_{N})),
\end{equation*}
where $p_{i}(s_{1},\cdots,s_{N})$ denotes the payment made to sensor $i$ which is in general a function of the strategies of all the sensors, and $\ell(s_{1},\cdots,s_{N})$ is the fusion rule used to obtain the global estimate. Note that the payment $p_i(s_1, s_2, \cdots, s_N)$ can also be expressed as $p_{i}(s_{i}, s_{-i})$. Denote the set of all feasible strategies $s_e$'s by $S_e$.
\item {\em Utility Functions:} The expected utility of each sensor $i$ is given by 
\begin{equation}
\expect{U_{i}} = \expect{p_{i}(s_{i}, s_{-i}) - c_{i}(\xi_{i})},
\end{equation}
where the expectation is taken over the uncertainties of the random variable $X$ and measurement noises. 
Thus each sensor $i$ optimizes over the effort level and reports to maximize its expected utility, 
\begin{equation}
\max_{s_{i} \in S_i} \\ \expect{U_i}.
\end{equation}
On the other hand, the estimator is interested in minimizing the expected total payment while obtaining a global estimate with MSE less than a certain threshold. Formally, the optimization problem at the estimator is given as follows
\begin{equation}   \label{optimization of center}
\begin{split}
\min_{s_e \in S_e} \quad &\expect*{\sum_{i=1}^{N} p_{i}(s_{i}, s_{-i})}\\
s.t. \quad &\Sigma_{g} \leq \Sigma_{t},\\
& \expect{p_{i}(s_{i}, s_{-i})-c_{i}(\xi_{i})}\geq 0, \forall i,\\
& s_i = \argmaxA\expect{p_{i}(s_{i}, s_{-i})- c_{i}(\xi_{i})} , \forall i,
\end{split}
\end{equation}
where $\Sigma_{t}$ is the specified threshold on the global MSE. The second constraint above ensures individual rationality, which is necessary for the sensors to participate.
\end{itemize}
In the sequel, we solve problem~(\ref{optimization of center}). Note that problem (\ref{optimization of center}) specifies a game among the sensors since their utilities depend on actions taken by all of them. We will consider the solution of the optimization problem when the behavior of the sensors is specified according to a Nash equilibrium.




\section{Optimization Problem at the Estimator}
\label{sec_ch2:optimization problem at the estimator}
To understand why the problem~(\ref{optimization of center}) is difficult to solve, we note why some intuitive incentive mechanisms may not work.
\begin{itemize}
    \item A payment scheme $p_{i} = c$ for a constant $c$ that does not depend on the reports will lead to each sensor not making any effort and reporting some arbitrary value to the estimator. In economics, this is termed as the problem of \textit{moral hazard}.
    \item A payment scheme that specifies $p_{i}$ as a decreasing function of $\Sigma_{i}$ or $\xi_{i}$ can be considered to incentivize the sensors to exert effort and take accurate measurements. However, it will lead sensors reporting very low local MSE irrespective of the actual effort made. This is termed as the problem of \textit{adverse selection}.
\end{itemize}
In either case, note that the actual measurements $y_i$, local estimates $\hat{x}_{i}$ and the local MSE $\Sigma_{i}$ are all unknown to the estimator, fusing reported local estimates to obtain a global estimate that satisfies the constraint $\Sigma_{g} \leq \Sigma_{t}$ is also a nontrivial problem. The overall optimization problem (\ref{optimization of center}) is even more difficult.

\begin{figure}[tpb] 
  \begin{center}
    \centerline{\includegraphics[width=\linewidth]{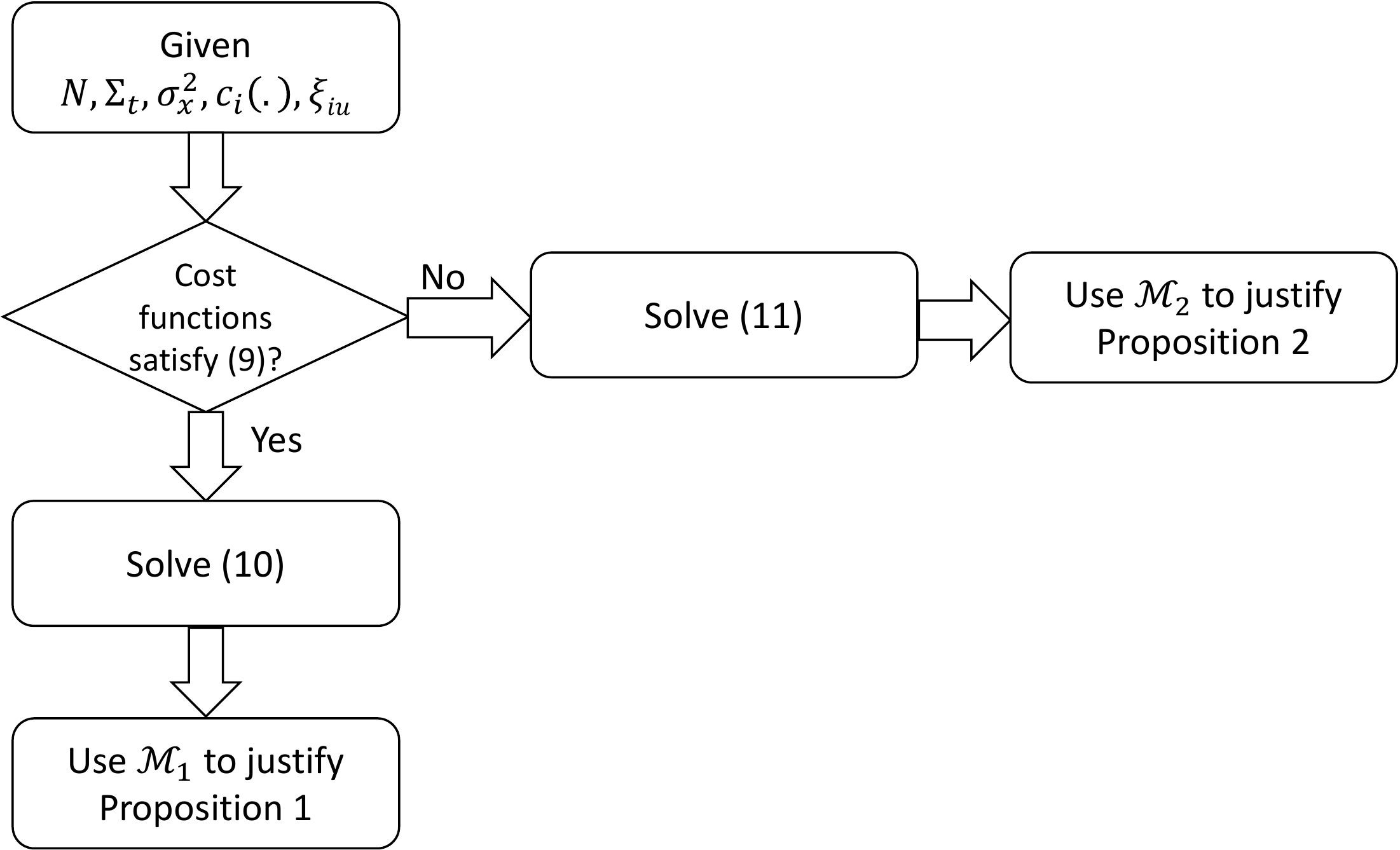}}
    \caption{The procedure of designing the incentive mechanism with strategic data sources.}
    \label{fig_ch2_flow_chart}
  \end{center}
\end{figure}
Our results are organized as shown in Fig. \ref{fig_ch2_flow_chart}. Specifically, we show that the following technical condition on the effort cost functions plays an important role in the  simplification of the problem:
\begin{equation} \label{constraint of realization}
    -2 \frac{\partial c_{i}(\xi_{i})}{\partial \xi_{i}}
    -
    \frac{\partial^{2} c_{i}(\xi_{i})}{\partial \xi_{i}^{2}}
    (\xi_{x}+{\xi}_{i})
    <0, \enskip \forall \xi_i \enskip \text{and} \enskip i.
\end{equation}

\begin{remark} \label{Remark:quadratic satisfy constraint}
If $c_{i}(\xi_{i})$ is convex over $\xi_{i}$, the constraint (\ref{constraint of realization}) holds for any $\xi_{i}$.
\end{remark}

Depending on whether~(\ref{constraint of realization}) is satisfied or not, we have the following result.
\begin{proposition} \label{proposition of using M1}
Consider the setup of problem~(\ref{optimization of center}). If condition~(\ref{constraint of realization}) is satisfied, the estimator can  specify a payment design such that
\begin{enumerate}
    \item the selected sensors exert the effort levels specified by the estimator;
    \item the selected sensors report truthfully about their measurements and local estimates;
    \item the expected payment to each selected sensor is the effort cost of the sensor for the specified effort level.
\end{enumerate}
\end{proposition}
Note that under these three conditions, the estimator can choose the optimal effort levels from agents that yield minimum payment while meeting the constraints in problem (\ref{optimization of center}) with the fusion rule as shown in (\ref{eq:fuse}). Thus, in this case, the optimization problem (\ref{optimization of center}) reduces to
\begin{equation} \label{eq:P3}
    \begin{split}
    \min_{\phi, \xi}  \quad & \sum_{i=1}^{N}\phi_{i}c_{i}(\xi_{i})\\
    s.t.  \quad &  \frac{1}{\xi_{x}+\sum_{i=1}^{N}\phi_{i}\xi_{i}}
    \leq \Sigma_{t},\\
    &  \phi_{i} \in (0,1),\\
    &  \xi_{i} \in [0,\xi_{iu}],
    \end{split}
\end{equation}
where $\xi = (\xi_1, \xi_2, \cdots, \xi_N)$ and $\phi = (\phi_{1}, \phi_{2}, \cdots, \phi_{N})$. $\phi_{i}$ is an indicator about whether or not the estimator selects agent $i$: $\phi_{i}=1$ represents the case where the estimator selects agent $i$ and $\phi_{i}=0$ 
represents the case where the estimator does not select agent $i$, which can be implemented by, for instance, setting $p_{i}=0$. 

In Section \ref{sec: M1}, we show that if the cost functions satisfy constraint (\ref{constraint of realization}), a mechanism $\M_1$ can be designed that specifies a payment design according to Proposition \ref{proposition of using M1}. Thus, problem (\ref{optimization of center}) can be solved optimally. 
Otherwise if the cost functions does not satisfy constraint (\ref{constraint of realization}), we solve the problem in a sub-optimal way through the following proposition. This will be proved through the design of a mechanism $\M_2$ presented in Section \ref{sub-optimal}.
\begin{proposition}
Consider the setup of problem~(\ref{optimization of center}). If the condition~(\ref{constraint of realization}) is not satisfied, the estimator can  specify a payment design such that
\begin{enumerate}
    \item the selected sensors would exert their maximum effort levels;
    \item the selected sensors report truthfully about their measurements and local estimates;
    \item the expected payment to each selected sensor $i$ is the effort cost of the sensor for its maximum effort level.
\end{enumerate}
\end{proposition}
Note that under these three conditions, the estimator cannot choose the optimal effort levels from sensors as in Proposition \ref{proposition of using M1}. 
But the estimator can still select a subset of all agents given that the selected agents will exert their maximum efforts and the estimator will pay the corresponding costs incurred at maximum effort levels. 
Formally, problem (\ref{optimization of center}) is transformed to the following well-defined binary knapsack problem (KP): 
\begin{equation} \label{eq:P7}
\begin{split}
\min_{{\phi}}  \quad &  \sum_{i=1}^{N}\phi_{i}c_{i}(\xi_{iu})\\
s.t. \quad &  \frac{1}{\xi_{x}+\sum_{i=1}^{N}\phi_{i}\xi_{iu}}
    \leq \Sigma_{t},\\
& \phi_{i} \in (0,1).
\end{split}
\end{equation}
This problem is NP-hard but can be solved exactly in pseudo-polynomial time through dynamic programming algorithms \cite{kellerer2003knapsack}.

\section{Optimal Mechanism When (\ref{constraint of realization}) Holds}       \label{sec: M1}

In this section, we address the cases where (\ref{constraint of realization}) holds. We first simplify the optimization problem (\ref{eq:P3}) and present two interesting special cases. Then we present an optimal mechanism to prove Proposition \ref{proposition of using M1}.

\subsection{Solving Problem (\ref{eq:P3})}
(\ref{eq:P3}) can be rewritten as
\begin{equation} \label{eq:rewrite P3}
    \begin{split}
    \min_{{\phi}, {\xi}} \quad & \sum_{i=1}^{N}\phi_{i}c_{i}(\xi_{i})\\
     s.t. \quad &
    \sum_{i=1}^{N}\phi_{i}\xi_{i}
    \geq \Sigma_{t}^{-1}-\xi_{x},\\
    &  \phi_{i} \in (0,1), \\
    &  \xi_{i} \in [0,\xi_{iu}].
    \end{split}
\end{equation}
This is a mixed-integer nonlinear programming problem and specifically known as the general knapsack problem (GKP) with variable coefficients \cite{suzuki1978generalized} \cite{holmberg1996solving}, which is difficult to solve in general. However, since $c_i(0)=0, ~ \forall i$, we can transform problem (\ref{eq:rewrite P3}) to problem (\ref{eq:P4}) according to the following result.

\begin{lemma}
Problem
(\ref{eq:rewrite P3}) can be solved by constructing solution of the following optimization problem
\begin{equation} \label{eq:P4}
    \begin{split}
    \min_{{\xi}}  \quad &  \sum_{i=1}^{N}c_{i}(\xi_{i})\\
    s.t.  \quad & 
    \sum_{i=1}^{N}\xi_{i}
    \geq \Sigma_{t}^{-1}-\xi_{x},\\
    & \xi_{i} \in [0,\xi_{iu}].
    \end{split}
\end{equation}
\end{lemma}

\begin{IEEEproof}
Suppose that the minimum of ($\ref{eq:rewrite P3}$), denoted by $O_{1}$, is achieved at $({\xi}^{O1}, {\phi}^{O1})$ and the minimum of ($\ref{eq:P4}$), denoted by $O_{2}$, is achieved at ${\xi}^{O2}$.  We have $O_{1}\leq O_{2}$ because ($\ref{eq:P4}$) is a special case of ($\ref{eq:rewrite P3}$) by fixing all $\phi_{i}=1$. On the other hand, $O_{1}\geq O_{2}$, because any value achieved in ($\ref{eq:rewrite P3}$) can be achieved in ($\ref{eq:P4}$) by constructing ${\xi}^{O2}$ from $({\xi}^{O1}, {\phi}^{O1})$ as 
\begin{equation}
    \xi^{O2}_{i} = 
    \begin{cases}
      \xi^{O1}_{i}, & \text{for } \phi^{O1}_{i}=1, \\
      0, & \text{for } \phi^{O1}_{i}=0. \\ 
    \end{cases}
\end{equation}
Thus, $O_{1}=O_{2}$. In general, it is easier to solve ($\ref{eq:P4}$) first and then construct $({\xi}^{O1}, {\phi}^{O1})$ from ${\xi}^{O2}$ by setting
\begin{equation}
    (\xi^{O1}_{i}, \phi^{O1}_{i}) = 
    \begin{cases}
      (\xi^{O2}_{i}, 1), & \text{for } \xi^{O2}_{i}\neq 0, \\
      (r, 0), & \text{for } \xi^{O2}_{i} = 0, \\ 
    \end{cases}
\end{equation}
where $r$ can be any number since $\phi^{O1}_{i}=0$.
\end{IEEEproof}

We now present two interesting special cases.
\subsubsection{Special Case: Continuous Quadratic Cost Function}
A quadratic effort cost $c_{i}(\xi_{i}) = l\xi_{i}^{2}$ is quite popular, e.g., in control theory. In this case,
then the optimization problem (\ref{eq:P4}) becomes,
\begin{equation} \label{eq:P6}
    \begin{split}
    \min_{{\xi}}  \quad & \sum_{i=1}^{N}l\xi^{2}_{i}\\
    s.t. \quad  & 
    \sum_{i=1}^{N}\xi_{i}
    \geq \Sigma_{t}^{-1}-\xi_{x},\\
    &  \xi_{i} \in [0,\xi_{iu}],
    \end{split}
\end{equation}
which is a standard Quadratic Programming (QP) problem. 
According to Cauchy-Schwarz inequality, the optimal solution of (\ref{eq:P6}) is given by
\begin{equation} \label{eq:P6 solution}
    \tilde{\xi}_{1} = \tilde{\xi}_{2} = ... = \tilde{\xi}_{N} = \frac{\Sigma_{t}^{-1}-\xi_{x}}{N},
\end{equation}
assuming for simplicity that $\frac{\Sigma_{t}^{-1}-\xi_{x}}{N}\leq\xi_{iu}$.

As stated in Remark \ref{Remark:quadratic satisfy constraint}, since the cost function is convex, the constraint (\ref{constraint of realization}) holds for any possible $\xi_{i}$. Under our optimal mechanism (presented in Section \ref{subsection: optimal mechanism}), there is a Nash equilibrium where all agents select the effort level as $\frac{\Sigma_{t}^{-1}-\xi_{x}}{N}$
and
report truthfully about their local estimates and their measurements. Meanwhile, the minimum expected total payment that can be achieved to ensure global MSE to be no greater than $\Sigma_{t}$ is $\dfrac{l(\Sigma_{t}^{-1}-\xi_{x})^{2}}{N}$.

\subsubsection{Special Case: Discrete Linear Cost Function} \label{special_case_discrete}
A natural model is that each agent can increase the accuracy of its local estimate by taking more measurements and estimating based upon the sample mean. For instance, if agent $i$ takes $\eta_{i}$ number of measurements of the following 
\begin{equation}
\begin{split}
&y_{i}(1) = x + v_{i}(1), \\
&y_{i}(2) = x + v_{i}(2), \\
&\vdots \\
&y_{i}(\eta_{i}) = x + v_{i}(\eta_{i}),
\end{split}
\end{equation}
where $v_{i}(k)$ follows i.i.d. Gaussian distribution $\mathcal{N} (0,\sigma^{2}_{io})$. Denote the effort cost of taking each measurement by a cost of $c_{io}$. 
Then the noise level, effort level and effort cost of the sample mean $\bar{y}_{i} =  x + \bar{v}_{i}$ averaged from taking $\eta_{i}$ measurements are respectively given by
\begin{equation}
\begin{split}
    &\sigma^{2}_{i} = \frac{\sigma^{2}_{io}}{\eta_{i}}, \\
    &\xi_{i}=\eta_{i}\sigma_{io}^{-2},\\
    &c_{i}(\xi_{i}) = \eta_{i}c_{io}=\sigma^{2}_{io}c_{io}\xi_{i}.
\end{split}
\end{equation}
Therefore in this case, the effort cost function a linear function and the effort level depends on the number of measurements taken.
We denote the corresponding maximum number of measurements by $\eta_{i}^{m}$.
The optimization problem (\ref{eq:P4}) becomes,
\begin{equation} \label{eq:P5}
    \begin{split}
    \min_{{\eta}}  \quad &  \sum_{i=1}^{N}\eta_{i}c_{io}\\
    s.t.  \quad  &
    \sum_{i=1}^{N}\eta_{i}\sigma^{-2}_{io}
    \geq \Sigma_{t}^{-1}-\xi_{x},\\
    & \eta_{i} \in (0, 1, ..., \eta^{m}_{i}),
    \end{split}
\end{equation}
which is known as the Bounded Knapsack Problem (BKP). It is NP-hard but it can be solved exactly in pseudo-polynomial time through dynamic programming algorithms \cite{kellerer2003knapsack}\cite{pisinger2000minimal}. Denote by $\tilde{{\eta}} = (\tilde{\eta}_{1}, \tilde{\eta}_{2}, ..., \tilde{\eta}_{N})$ the optimal solution of (\ref{eq:P5}).
%

Under our optimal mechanism, there is a Nash equilibrium where each agent takes $\tilde{\eta}_{i}$ number of measurements and report truthfully about the local estimate and measurement. Meanwhile, the minimum expected total payment that can be achieved to ensure global MSE to be no greater than $\Sigma_{t}$ is  $\sum_{i=1}^{N}\tilde{\eta}_{i}c_{io}$.

\subsection{Optimal Mechanism} \label{subsection: optimal mechanism}
Denoting the optimal solution of problem (\ref{eq:P4}) by $\tilde{{\xi}} = (\tilde{\xi}_1, \tilde{\xi}_2, ..., \tilde{\xi}_N)$, we now present mechanism $\M_{1}$ that fulfills Proposition \ref{proposition of using M1}, i.e., under Mechanism $\M_{1}$, all the agents exerting the desired effort levels and reporting their measurements and local estimates truthfully is a Nash equilibrium. In addition, the expected payment to each agent $i$ is the effort cost of the agent for the specified effort level.

In our proposed incentive mechanism $\M_{1}$, agents are asked to report two items $(\hat{x}_{ri}, y_{ri})$, where $\hat{x}_{ri}$ is the reported local estimate and $y_{ri}$ is the reported measurement. Note that $\hat{x}_{ri} \neq \hat{x}_{i}$ and $y_{ri} \neq y_{i}$ in general since agents may falsify their reports to maximize their utilities.
The payment function is given by
\begin{equation} \label{p2}
p_i(\hat{x}_{ri},y_{rj})=
  \gamma_i-\beta_i(\hat{x}_{ri}-y_{rj})^2,
\end{equation}
where $y_{rj}$ is the reported measurement from another agent $j \neq i$.
%
%
%
%
As before, agents are interested in maximizing their expected utilities,
\begin{equation}
s^{*}_{i} = \argmaxA_{s_{i} \in S_i}\E\left[U_{i}\right]=\argmaxA_{s_{i} \in S_i}\E\left[p_i(\hat{x}_{ri},y_{rj})-c_{i}(\xi_{i})\right].
\end{equation}
%

Now, we state our results about the optimal mechanism $\M_{1}$.
\begin{theorem} \label{Theorem 1}
Consider the problem (\ref{optimization of center}) when (\ref{constraint of realization}) holds. Let (\ref{p2}) be the payment function to each sensor $i$ with
\begin{equation} \label{beta}
\beta_{i}=\evalat[\bigg]{\frac{\partial c_{i}(\xi_{i})}{\partial \xi_{i}}}{\xi_{i}=\tilde{\xi}_i}(\xi_{x}+\tilde{\xi}_{i})^2,
\end{equation}
and 
\begin{equation} \label{gamma}
    \gamma_{i}=\beta_{i}\left(\frac{1}{\xi_{x}+\tilde{\xi}_{i}}+\tilde{\xi}_{j}^{-1}\right)+c_{i}(\tilde{\xi}_{i}).
\end{equation}
The strategy profile
$s^{*}=(s_{1}^{*},s_{2}^{*},...,s_{N}^{*})$ with 
\begin{equation}
s^{*}_{i} = 
  (\xi_{i}=\tilde{\xi}_{i}, \hat{x}_{ri}=\hat{x}_{i}, y_{ri}=y_{i})
\end{equation}
is a Nash equilibrium of the mechanism design problem (\ref{optimization of center}).
In addition, the expected payment to each agent is the effort cost, i.e., $\expect{p_{i}}=c_{i}(\tilde{\xi}_{i})$.
\end{theorem}
      
\begin{IEEEproof}
See Appendix \ref{apdx:ch2_th1}.
\end{IEEEproof}

Intuitively, the payment is designed as a function of the difference between the reports from the agents, which motivates each agent to estimate the information of another agent based on her own information. The accuracy of the agent's estimate, and the corresponding expected payment, will depend on how much effort is exerted when the agent obtains her own information. Therefore, $\beta_{i}$ can be designed as in (\ref{beta}) so that the effort level that the estimator wishes each agent to exert (i.e., $\tilde{\xi_{i}}$) will turn out to be exactly the optimal choice for agent $i$ when all the other agents exert the effort levels desired by the estimator, i.e., $\xi_{j} = \tilde{\xi}_j, \forall j\neq i$. Further, $\gamma_{i}$ can be designed as in (\ref{gamma}) so that the expected payment is small but enough to cover the effort cost. 

It is worth remarking that the estimation of the reports among the agents is in a \textit{game} setting, which means the desired strategy profile $s^{*}$ from all the agents is obtained in a Nash equilibrium sense.
Further, if there exists an \textit{`honest'} agent who reports its measurement and effort level truthfully, it is no longer needed to ask the strategic agents to report their measurements. In this case, the desired strategy profile $s^{*}$ is an {\it equilibrium in strictly dominant strategies} in which every $s_i^*$ is the strictly dominant strategy for agent $i$. We present the result in the following corollary.   

\begin{corollary} \label{corollary:unique NE}
Consider the setting of Theorem \ref{Theorem 1} with an honest agent $h$ who reports its measurement and effort level truthfully, i.e., $y_{rh}=y_{h}=x+v_h$, where $v_h\sim\mathcal{N} (0, \xi_{h}^{-1})$. Let the payment function to each sensor $i$ be specified by (\ref{p2}), (\ref{beta}) and (\ref{gamma}) after replacing $y_{rj}$ and $\tilde{\xi}_j$ with $y_{rh}$ and ${\xi}_h$ respectively.
The strategy profile $s^{*'}=(s_{1}^{*'},s_{2}^{*'},...,s_{N}^{*'})$ with 
\begin{equation}
s^{*'}_{i} = 
  (\xi_{i}=\tilde{\xi}_{i}, \hat{x}_{ri}=\hat{x}_{i})
\end{equation}
is the unique equilibrium in strictly dominant strategies.
\end{corollary}
\begin{IEEEproof}
The proof is similar to the proof of Theorem \ref{Theorem 1}. The only difference in this case is that the strategic agents now estimate the measurement from the honest agent instead of estimating the measurement from another strategic agent. The utility of each strategic agent will no longer depend on other strategic agents, thus, the strategy in the Nash equilibrium is the strictly dominant strategy for each agent and the Nash equilibrium is the unique equilibrium. 
\end{IEEEproof}

\section{A Sub-optimal Mechanism When (\ref{constraint of realization}) Does Not Hold} \label{sub-optimal}
In this section, we provide a feasibility-guaranteed sub-optimal mechanism $\M_{2}$ for the cases where the constraint (\ref{constraint of realization}) can not be satisfied. $\M_{2}$ achieves truthful reporting and elicits maximum effort from the selected agents with expected payment to selected agent $i$ being the effort cost $c_{i}(\xi_{iu})$ in a Nash equilibrium sense.

Denote the optimal solution to (\ref{eq:P7}) as $\tilde{{\phi}} = (\tilde{\phi}_{1}, \tilde{\phi}_{2}, ..., \tilde{\phi}_{N})$. We present the incentive mechanism $\M_{2}$ that only selects the agents for which $\tilde{\phi}_{i}=1$ and elicits their maximum efforts.  

\begin{theorem} \label{Theorem 2}
Consider the problem (\ref{optimization of center}) when (\ref{constraint of realization}) does not hold. Let the payment to each sensor $i$ with $\tilde{\phi}_{i}=1$ be determined by comparing its reported local estimate with the reported measurement from another agent $j$ with $\tilde{\phi}_{j}=1$, i.e.,
\begin{equation} \label{payment theorem 2}
p_i(\hat{x}_{ri},y_{rj})=
\begin{cases}
  \gamma_i-\beta_i(\hat{x}_{ri}-y_{rj})^2,
  & \text{for } \tilde{\phi}_{i}=1 \\
  0, 
  & \text{for } \tilde{\phi}_{i}=0
\end{cases}
\end{equation}
with 
\begin{equation} \label{beta theorem 2}
\beta_{i}>\max_{\xi_{i}\in [0,\xi_{iu}]}{\frac{\partial c_{i}(\xi_{i})}{\partial \xi_{i}}}(\xi_{x}+\xi_{i})^{2}
\end{equation}
and 
\begin{equation} \label{gamma theorem 2}
    \gamma_{i}=\beta_{i}(\frac{1}{\xi_{x}+\xi_{iu}}+\xi_{ju}^{-1})-c_{i}(\xi_{iu}).
\end{equation}
The strategy profile ${s}^{*}=(s_{1}^{*}, s_{2}^{*}, ..., s_{N}^{*})$ with 
\begin{equation}
s^{*}_{i} = 
\begin{cases}
  (\xi_{i}=\xi_{iu}, \hat{x}_{ri}=\hat{x}_{i}, y_{ri}=y_{i}), 
  & \text{for } \tilde{\phi}_{i}=1 \\
  (\xi_{i}=0, \hat{x}_{ri}=\hat{x}_{i}, y_{ri}=y_{i}), 
  & \text{for }  \tilde{\phi}_{i}=0
\end{cases}
\end{equation}
is a Nash equilibrium of the mechanism design problem (\ref{optimization of center}).
In addition, the expected payment to each agent is the effort cost for its maximum effort level, i.e., $\expect{p_{i}}=c_{i}(\xi_{iu})$.
\end{theorem}

\begin{IEEEproof}
See Appendix \ref{apdx:ch2_th2}.
\end{IEEEproof}

Similarly, if there exists an honest agent who reports its measurement and effort level truthfully, it is no longer needed to ask the strategic agents to report their measurements. In this case, the desired strategy profile is the unique equilibrium with strictly dominant strategies.
\begin{corollary}
Consider the setting in Theorem \ref{Theorem 2} with an honest agent $h$ who reports its measurement and effort level truthfully, i.e., $y_{rh}=y_{h}=x+v_h$, where $v_h\sim\mathcal{N} (0, \xi_{h}^{-1})$. Let the payment function be specified by (\ref{payment theorem 2}), (\ref{beta theorem 2}) and (\ref{gamma theorem 2}) after replacing $y_{rj}$ and $\tilde{\xi}_j$ with $y_{rh}$ and ${\xi}_h$ respectively.
The strategy profile ${s}^{*'}=(s_{1}^{*'},s_{2}^{*'},...,s_{N}^{*'})$ with 
\begin{equation}
s^{*'}_{i} = 
\begin{cases}
  (\xi_{i}=\xi_{iu}, \hat{x}_{ri}=\hat{x}_{i}), 
  & \text{for } \tilde{\phi}_{i}=1 \\
  (\xi_{i}=0, \hat{x}_{ri}=\hat{x}_{i}), 
  & \text{for }  \tilde{\phi}_{i}=0
\end{cases}
\end{equation}
is the unique equilibrium in strictly dominant strategies.
\end{corollary}

The proof is omitted since it is similar to that of Corollary \ref{corollary:unique NE}.

\section{Simulation Experiments}

\begin{figure}[tpb]
  \begin{center}
    \centerline{\includegraphics[width=\linewidth]{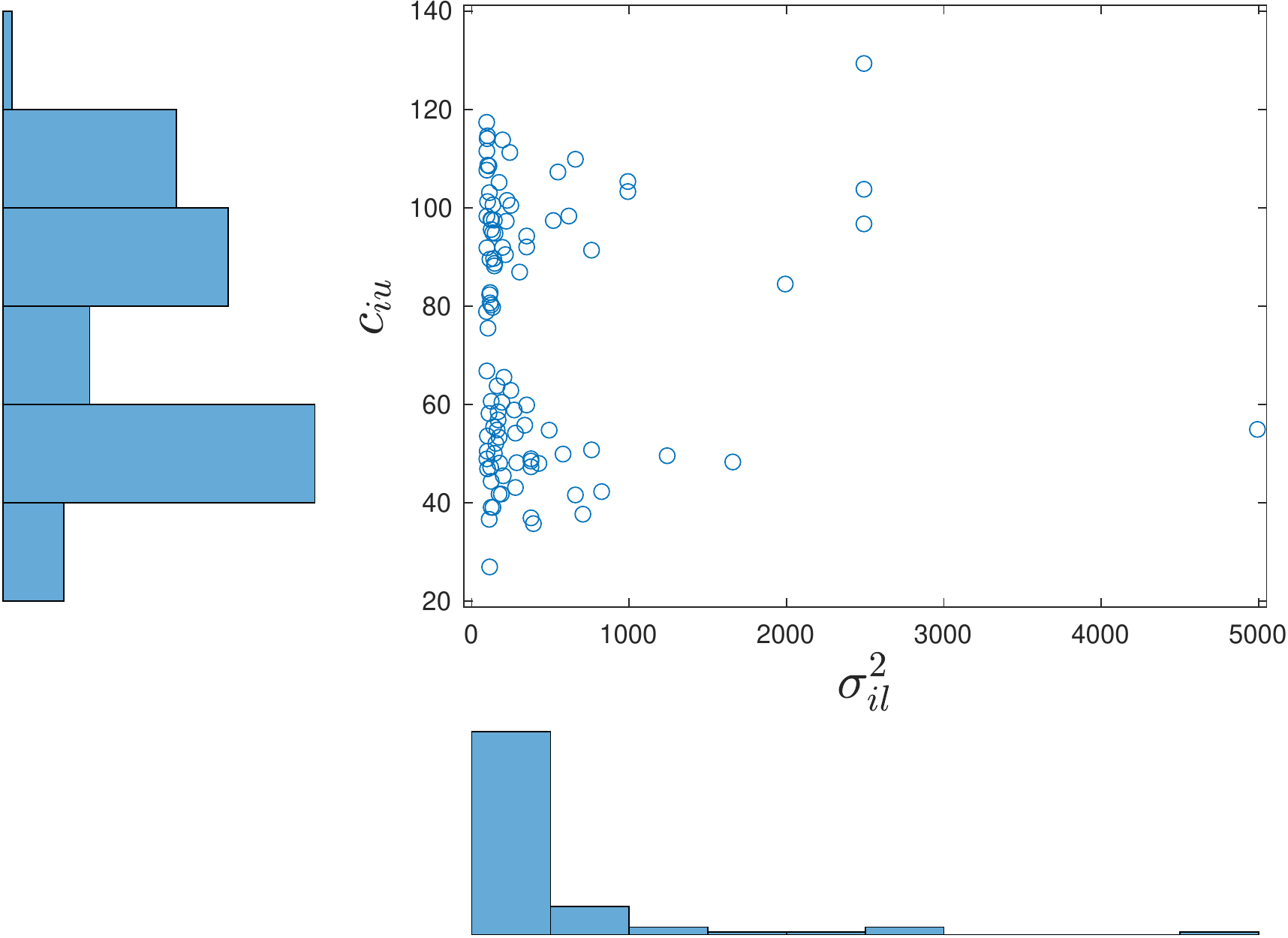}}
    \caption{Scatter plot and histograms of $\sigma^{2}_{il}$ and $c_{iu}$.}
    \label{fig:distribution of parameters}
  \end{center}
\end{figure}

\begin{figure}[tpb]
  \begin{center}
    \centerline{\includegraphics[width=\linewidth]{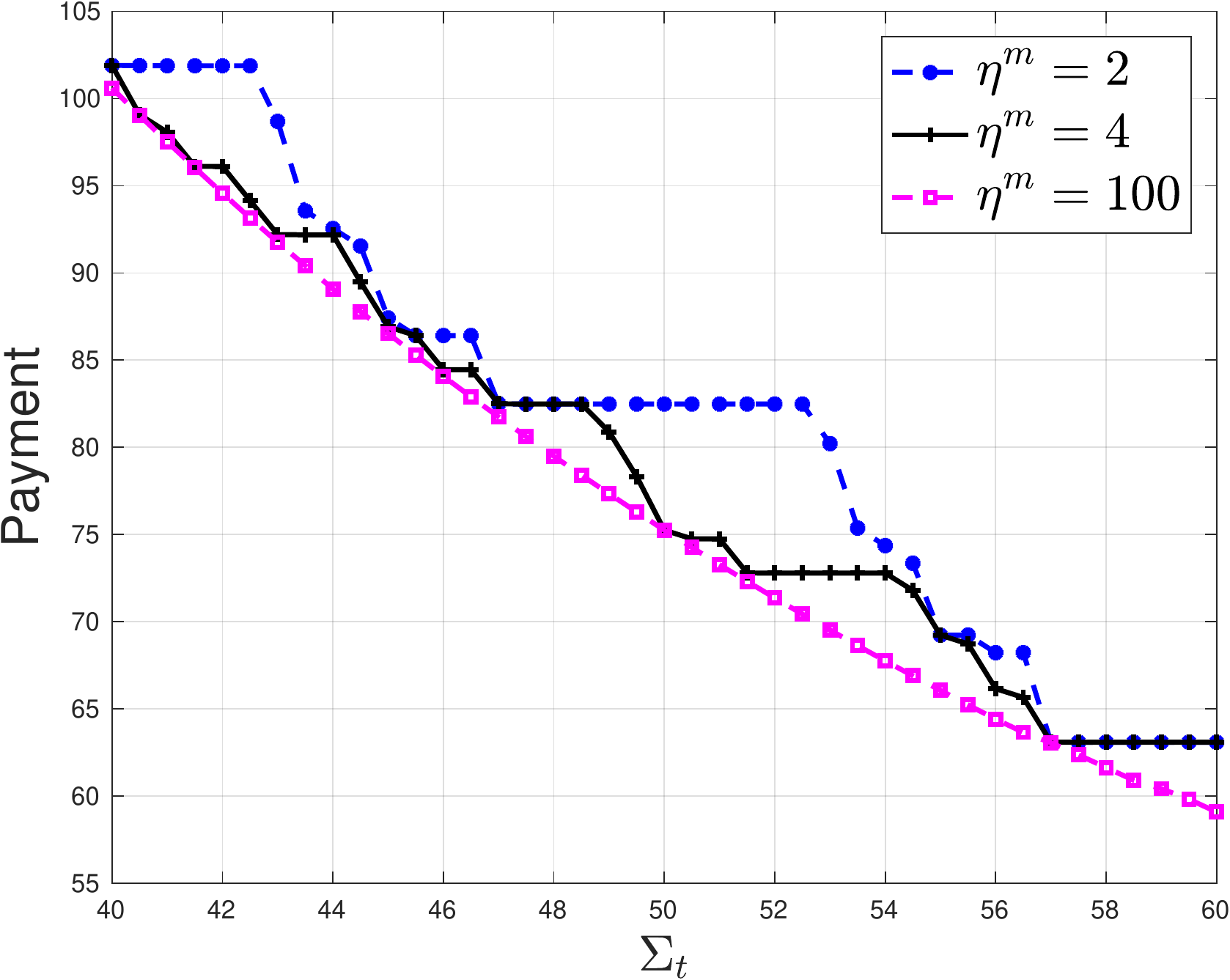}}
    \caption{Minimum payments with $N=100$, and $\eta^{m}=2$, $\eta^{m}=4$, $\eta^{m}=100$ respectively.}
    \label{fig:different eta_m}
  \end{center}
\end{figure}

\begin{figure}[tpb]
  \begin{center}
    \centerline{\includegraphics[width=\linewidth]{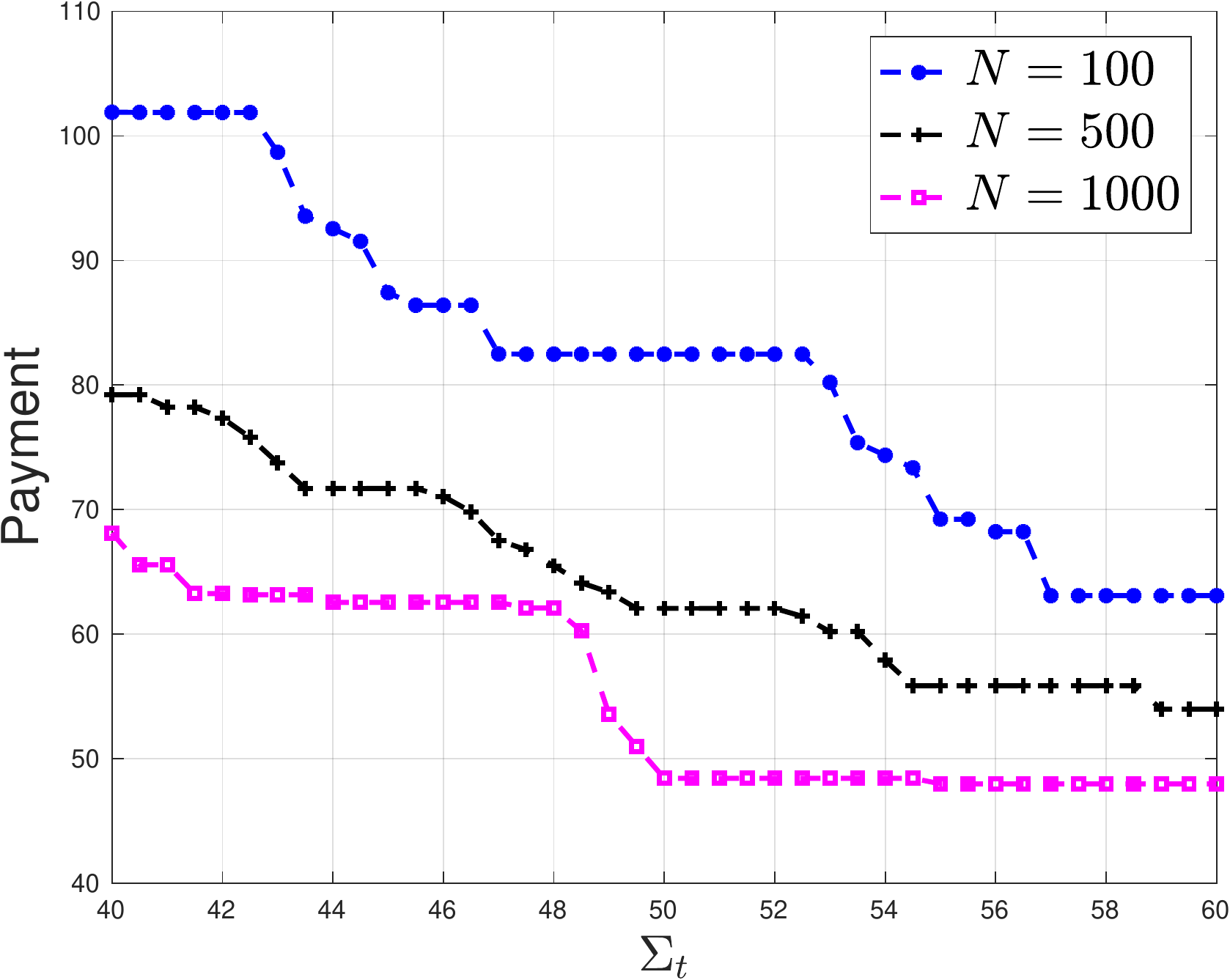}}
    \caption{Minimum payments with $\eta^{m}=2$, and $N=100$, $N=500$, $N=1000$ respectively.}
    \label{fig: comparison_N}
  \end{center}
\end{figure}

\begin{figure}[tpb]
  \begin{center}
    \centerline{\includegraphics[width=\linewidth]{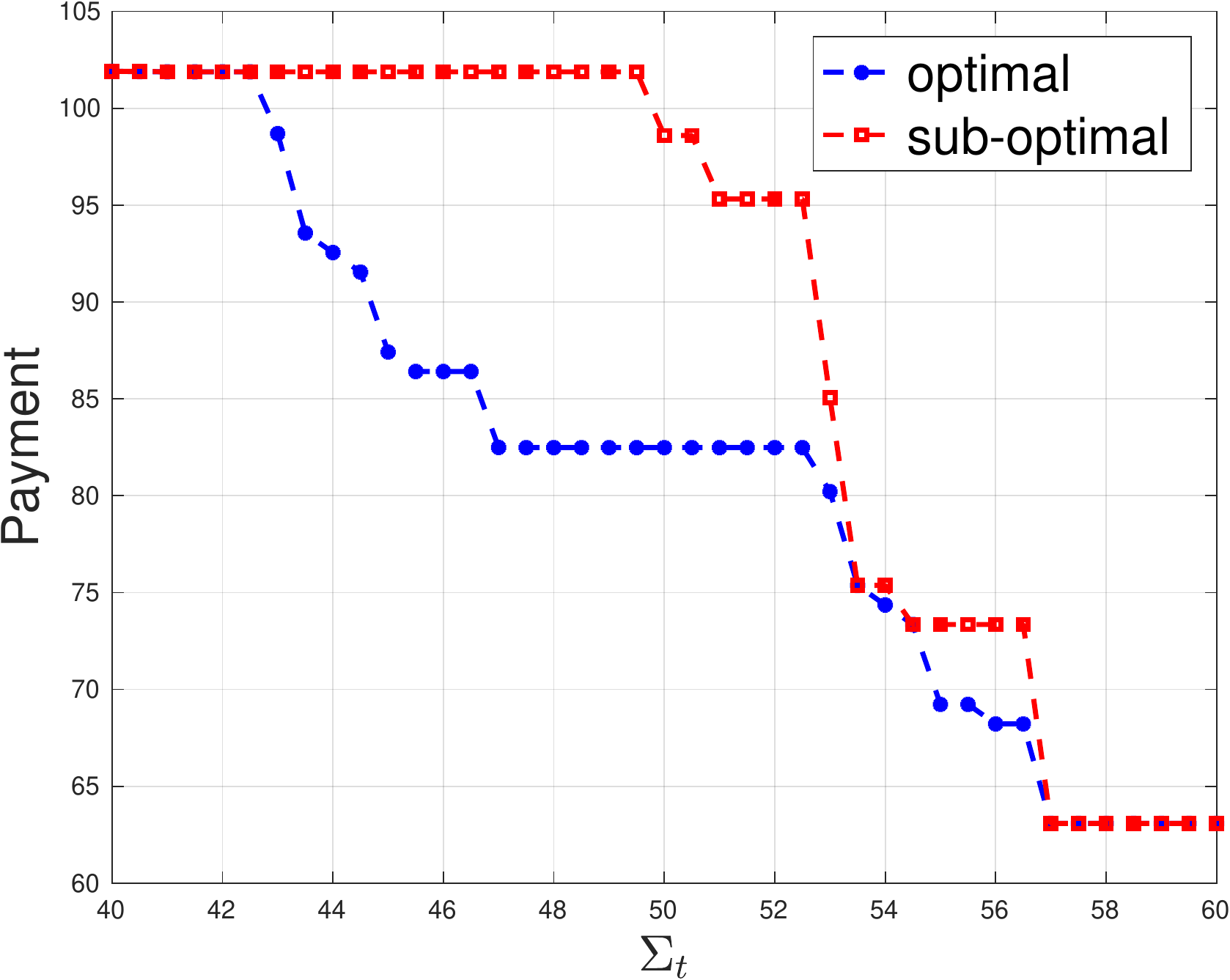}}
    \caption{Comparison of minimum payments in the sub-optimal case and the optimal case.} \label{fig:compare_suboptimal}
  \end{center}
\end{figure}

In this section, we demonstrate our mechanisms with simulation experiments. We first consider the setting of the problem in Section \ref{special_case_discrete} and investigate the minimum payment at different threshold $\Sigma_{t}$. $N=100$ agents are simulated and the variance of the prior distribution is selected as $\sigma^{2}_{x}=1000$. Further, fixing the minimum variance of each agent $\sigma^{2}_{il}$ and its corresponding maximum effort $c_{iu}$ allows us to study the effect $\eta^{m}_{i}$, which can be interpreted as the quantization level of the effort cost of each agent. 
Without loss of generosity, we set $\eta^{m}_{i}=\eta^{m}$ for all $i$.
To make the agents heterogeneous on their highest accuracies, we randomly generate $\sigma^{-2}_{il}\sim\mathcal{U}[0.0001, 0.01]$, which is selected such that roughly a half of $\sigma^{2}_{il}$ fall in the range $[100, 200]$ and the other half of $\sigma^{2}_{il}$ fall in the range $[200, 10000]$. $c_{iu}$ is randomly generated from a mixture Gaussian distribution $c_{iu}\sim.5\mathcal{N}(50, 100)+.5\mathcal{N}(100, 100)$.
The scatter plot and histograms of these two parameters are shown in Fig. \ref{fig:distribution of parameters}.

The minimum payments with $\eta^{m}=2$, $\eta^{m}=4$, and $\eta^{m}=100$ at different threshold $\Sigma_{t}$ are shown in Fig. \ref{fig:different eta_m}. In general, greater $\eta^{m}$ yields smaller payment. 
On the other hand, we also study the effect of $N$. We use the same distributions to generate $\sigma^{2}_{il}$ and $c_{iu}$. $\eta^{m}$ is fixed as $\eta^{m}=2$. As shown in Fig. \ref{fig: comparison_N}, more agents being available generally yields smaller payments.
Lastly, we compare our sub-optimal case with the optimal case considered in Section \ref{special_case_discrete} under the same setting. Recall that in the sub-optimal solution, our mechanism $\M_{2}$ yields all selected agents exerting maximum effort.
Using the same simulated parameters, the optimization problem (\ref{eq:P7}) in the sub-optimal case can be viewed a problem similar to (\ref{eq:P5}), but the decision variables are limited to be either $0$ or $\eta^{m}$. 
In Fig. \ref{fig:compare_suboptimal}, we show the comparison of minimum payments between the sub-optimal case and the optimal case at different $\Sigma_{t}$ with $\eta^{m}=2$ and $N=100$.

\section{Summary}
In this paper, we designed incentive mechanisms for a static estimation problem where the data sources are strategic agents whose measurements and accuracies are both unknown to the estimator. The objective of the incentive mechanism is to minimize the expected total payment made to the agents with a guaranteed quality of global estimate. We formulate the problem in a very general setting without assuming any specific format of the agents' cost functions. Instead, we designed an optimal incentive mechanism for the cases where the cost functions satisfy certain property and provided a sub-optimal incentive mechanism for the other cases. We also demonstrated our mechanisms by two special cases with continuous quadratic cost function and discrete linear cost function. Both in the special case with the discrete linear cost function and in the sub-optimal case, the optimization problem were transformed to knapsack problems, which can be solved in pseudo-polynomial time by dynamic programming.
Future work will include extending the results to dynamic estimation problems.

\appendices
\section{Proof of Theorem \ref{Theorem 1}} \label{apdx:ch2_th1}

It suffices to prove that if the strategy profile of all the other agents follow the stated equilibrium, denoted as ${s}_{-i}={s}_{-i}^{*}$, agent $i$ does not have another strategy which yields greater expected utility than ${s}_{i}^{*}$. Mathematically, when ${s}_{-i}={s}_{-i}^{*}$, the optimal strategy for agent $i$ is given by 
\begin{equation}
\begin{split}
  (\xi^{*}_{i}, \hat{x}^{*}_{ri}, y^{*}_{ri}) 
  &=  \argmaxA \expect{U_{i}} \\
  &=  \argmaxA \expect*{\gamma_i-\beta_i(\hat{x}_{ri}-y_{j})^2 - c_{i}(\xi_{i})|\hat{x}_{i}}. \\
\end{split}
\end{equation}

First notice that the estimator can not verify the reports $\hat{x}^{*}_{ri}$ and $y^{*}_{ri}$ jointly, since
\begin{equation}
    \hat{x}_{i} = \frac{\xi_{x}^{-1}}{\xi_{x}^{-1}+\xi_{i}^{-1}}y_{i},
\end{equation}
and $\xi_{i}$ is unknown to the estimator. Therefore, the agent can optimize $\hat{x}^{*}_{ri}$ and $y^{*}_{ri}$ independently. However, the expected utility is indifferent to $y_{ri}$, hence no other value can yield greater utility than $y^{*}_{ri}=y_{i}$.

Next, we prove that for any given $\xi_{i}$, the optimal $\hat{x}^{*}_{ri}=\hat{x}_{i}$. Since $\beta_{i}$ and $\gamma_{i}$ are positive constants and $\xi_{i}$ is fixed,
\begin{equation}
\begin{split}
    \hat{x}^{*}_{ri} &= \argminA \expect*{(\hat{x}_{ri}-y_{j})^{2}|\hat{x}_{i}} \\
    &= \expect*{y_{j}|\hat{x}_{i}} \\
    &= C_{y_{j}\hat{x}_{i}}C_{\hat{x}_{i}\hat{x}_{i}}^{-1}\hat{x}_{i}, \\
\end{split}
\end{equation}
where $C_{y_{j}\hat{x}_{i}}$ and $C_{\hat{x}_{i}\hat{x}_{i}}$ are computed as 
\begin{equation}
     C_{y_{j}\hat{x}_{i}} 
     = \expect*{(x+v_{j})\frac{\xi_{x}^{-1}}{\xi_{x}^{-1}+\xi_{i}^{-1}}(x+v_{i})}
     = \frac{\xi_{x}^{-2}}{\xi_{x}^{-1}+\xi_{i}^{-1}},
\end{equation}
and
\begin{equation}
     C_{\hat{x}_{i}\hat{x}_{i}} 
     = \expect*{\left(\frac{\xi_{x}^{-1}}{\xi_{x}^{-1}+\xi_{i}^{-1}}(x+v_{i})\right)^{2}} 
     = \frac{\xi_{x}^{-2}}{\xi_{x}^{-1}+\xi_{i}^{-1}},
\end{equation}
since the noises and $x$ are all independent to each other.

Now, we show that with $\beta_{i}$ given by (\ref{beta}) and if the constraint (\ref{constraint of realization}) is satisfied at $\tilde{\xi}_{i}$, then the optimal $\xi^{*}_{i}=\tilde{\xi}_{i}$. For any given $\xi_{i}$, the expected utility with the optimal choice of $\hat{x}_{ri}$ is given by 
\begin{equation}
    \expect{U_{i}(\xi_{i})} = \gamma_{i} - \beta_{i}\expect{(\hat{x}_{i}-y_{j})^{2}} - c_{i}(\xi_{i}),
\end{equation}
where 
\begin{equation}
\begin{split}
    \expect{(\hat{x}_{i}-y_{j})^{2}} 
    &= \expect*{\left(\frac{\xi_{x}^{-1}}{\xi_{x}^{-1}+\xi_{i}^{-1}}(x+v_{i})-(x+v_{j})\right)^{2}} \\
    &= \frac{\xi_{i}^{-1}\xi_{x}^{-1}}{\xi_{i}^{-1}+\xi_{x}^{-1}} + \xi_{j}^{-1} \\
    & = \frac{1}{\xi_{i}+\xi_{x}} + \tilde{\xi}_{j}^{-1}.
\end{split}
\end{equation}
Thus, setting the first derivative of $\expect{U_{i}(\xi_{i})}$ over $\xi_{i}$ to zero yields the unique maximum $\xi^{*}_{i}$ if $\expect{U_{i}(\xi_{i})}$ is concave:
\begin{equation} \label{eq:derivative over y_i}
    \frac{\partial\expect{U_{i}(\xi_{i})}}{\partial \xi_{i}} 
    = \frac{\beta_{i}}{(\xi_{i}+\xi_{x})^{2}}
    -
    \frac{\partial c_{i}(\xi_{i})}{\partial \xi_{i}} = 0,
\end{equation}
where the solution is $\tilde{\xi}_{i}$ if $\beta_{i}$ is given by
\begin{equation}
\beta_{i}=\evalat[\bigg]{\frac{\partial c_{i}(\xi_{i})}{\partial \xi_{i}}}{\xi_{i}=\tilde{\xi}_{i}}(\tilde{\xi}_{i}+\xi_{x})^2.
\end{equation}
To guarantee the concavity of $\expect{U_{i}(\xi_{i})}$, 
\begin{equation}
    \frac{\partial^{2}\expect{U_{i}(\xi_{i})}}{\partial \xi_{i}^{2}} 
    = \frac{-2\beta_{i}}{(\xi_{i}+\xi_{x})^{3}}
    -
    \frac{\partial^{2} c_{i}(\xi_{i})}{\partial \xi_{i}^{2}}<0,
\end{equation}
which implies the constraint (\ref{constraint of realization}) should be satisfied for any $\xi_{i}$.

Lastly, the maximum expected utility of agent $i$ is given by 
\begin{equation}
    \expect{\tilde{U}_{i}} = \gamma_{i} - \beta_{i}\left(\frac{1}{\xi_{x}+\tilde{\xi}_{i}}+\tilde{\xi}_{j}^{-1}\right) - c_{i}(\tilde{\xi}_{i}). 
\end{equation}
Therefore, $\gamma_{i}$ given by (\ref{gamma}) is designed to satisfy individual rationality. Meanwhile, the expected payment is as small as the effort cost $c_{i}(\tilde{\xi}_{i})$.

\section{Proof of Theorem \ref{Theorem 2}} \label{apdx:ch2_th2}
The proof is similar to the proof of Theorem \ref{Theorem 1}, except for that $\beta_{i}$ is designed to ensure that the derivative over $\xi_{i}$ as shown in (\ref{eq:derivative over y_i}) is always positive so that the selected agents would prefer to exert maximum effort.

\bibliographystyle{IEEEtran}
\bibliography{bibliography}

\end{document}